# Superconductivity close to magnetic instability in Fe(Se$_{1-x}$Te$_x$)$_{0.82}$

M.H. Fang[1], H.M. Pham[2], B. Qian[1], T.J. Liu[1], E. K. Vehstedt[1], Y. Liu[3], L. Spinu[2], and Z. Q. Mao[1]

[1] Department of Physics, Tulane University, New Orleans, Louisiana 70118, USA

[2] Advanced Materials Research Institute and Department of Physics, University of New Orleans, New Orleans, Louisiana 70148, USA

[3] Department of Physics, The Pennsylvania State University, University Park, Pennsylvania 16801, USA

Abstract

We report our study of the evolution of superconductivity and the phase diagram of the ternary Fe(Se$_{1-x}$Te$_x$)$_{0.82}$ ($0 \leq x \leq 1.0$) system. We discovered a new superconducting phase with $T_{c,max}$ = 14 K in the $0.3 < x < 1.0$ range. This superconducting phase is suppressed when the sample composition approaches the end member FeTe$_{0.82}$, which exhibits an incommensurate antiferromagnetic order. We discuss the relationship between the superconductivity and magnetism of this material system in terms of recent results from neutron scattering measurements. Our results and analyses suggest that superconductivity in this new class of Fe-based compounds is associated with magnetic fluctuations, and therefore may be unconventional in nature.

PACS numbers: 74.70.-b, 74.25.Dw, 74.25.Fy

I. INTRODUCTION

The discovery of high temperature superconductivity up to 56 K in the iron arsenide compounds $LnO_{1-x}F_xFeAs$ (Ln = Lanthanides) [1-6] is quite surprising since iron ions in many compounds have magnetic moments and they normally form an ordered magnetic state. Neutron scattering investigations of these materials have demonstrated that there exists a long-range spin density wave (SDW) type antiferromagnetic order in the undoped parent compound LaOFeAs [7, 8]. This suggests that magnetic fluctuations may play an essential role in mediating superconducting pairing in doped materials [9-11], similar to the scenario seen in high-$T_c$ cuprates. The newly-discovered binary superconductor FeSe ($T_c \approx$ 10 K) is another example of iron-based superconductor [12]. Interestingly, this binary system contains antifluorite planes which are isostructural to the FeAs layer found in the quaternary iron arsenide [13]. The $T_c$ of this material was increased to 27 K by applying hydrostatic pressure [14], suggesting that the simple binary FeSe may possess some essential ingredients for achieving high temperature superconductivity in FeAs-based compounds. Band structure calculations show that the Fermi surface structure of FeSe is indeed very similar to that of the FeAs-based compounds [15].

FeSe has a complicated phase diagram originating from nonstoimetric compositions [16]. The structure and magnetic properties of this system depend sensitively on the relative ratio of Se/Fe. For example, $FeSe_{0.82}$ has a PbO-type structure with a tetragonal space group *P4/nmm* and is superconducting, while $FeSe_{1.14}$ has a hexagonal structure and is a ferrimagnet [16]. In order to examine if the superconductivity in FeSe is associated with magnetism, the magnetic properties of other related iron chalcogen binary compounds possessing a layered tetragonal structure similar to FeSe should be examined. We note that in the FeTe binary system the composition in the $FeTe_{0.85}$-$FeTe_{0.95}$ range is tetragonal, isostructural to the $FeSe_{0.82}$ superconductor [12] and ferrimagnetic [17]. Given that $FeSe_{0.82}$ is superconducting, it is particularly interesting to investigate how the superconducting state evolves toward a magnetically ordered state with Te substitution for Se. For this purpose, we prepared polycrystalline samples of the $Fe(Se_{1-x}Te_x)_{0.82}$ ($0 \leq x \leq 1.0$) series and characterized their electronic and magnetic properties. We discovered two different superconducting phases, one for $0 \leq x < 0.15$ and the other for $0.3 < x < 1.0$, and the coexistence of the two phases for $0.15 \leq x \leq 0.3$. The $0.3 < x < 1.0$ phase has the highest superconducting transition temperature of $T_{c,max}$ = 14 K under ambient pressure. Most importantly, we found that this superconducting phase is suppressed only when the sample composition approaches the end member $FeTe_{0.82}$, which has a long-range magnetic order. These findings strongly suggest that superconductivity in Fe-based compounds is associated with magnetic fluctuations, and therefore may be unconventional in nature.

II. EXPERIMENT

Our samples with nominal compositions $Fe(Se_{1-x}Te_x)_{0.82}$ (x=0, 0.05, 0.2, 0.25, 0.3, 0.4, 0.5, 0.6, 0.7, 0.8, 0.9, and 1.0) were prepared using a solid state reaction method. The mixed powder was first pressed into pellets, then sealed in an evacuated quartz tube and sintered at 700 °C for 24 hours. The sample was then reground, pressed into pellets, and sintered again at 700 °C for 24 hours. Structural characterization of these samples was performed using x-ray diffractometery. Resistivity measurements were performed using a standard four-probe method in a Physical Property Measurement System (PPMS, Quantum

Design). The magnetization was measured using a SQUID (Quantum Design). Hall effects for the samples with $x = 0.6$ and 1.0 were measured using conventional four-probe method; the longitudinal resistivity component was eliminated by reversing the field direction.

III. RESRULTS AND DISCUSSIONS

X-ray diffraction analyses showed all of our samples to be of high purity. Only a negligible amount of impurity phase, β-FeSe, was observed in the samples near the Se side. Figure 1 shows x-ray diffraction patterns of typical compositions. We find that the diffraction peaks of both end members, $FeSe_{0.82}$ and $FeTe_{0.82}$, can be indexed with the tetragonal lattice *P4/nmm*, which is consistent with previously reported results [12, 17], but their lattice parameters are remarkably different. Diffraction peaks exhibit systematic shifts with the variation of the Se/Te ratio either for $x > 0.3$ or $x < 0.15$. For $0.15 < x < 0.3$, however, all diffraction peaks split into two sets, implying a coexistence of two structural phases (e.g., the data of the $x = 0.25$ sample in Fig. 1). This observation suggests that the structure of $FeTe_{0.82}$ is essentially different from that of $FeSe_{0.82}$ even though both of them can be described by a similar tetragonal lattice. Here we use A and B to denote these two structure phases respectively. Structure A is stable for $0 \leq x < 0.15$, while Structure B is stable for $0.3 < x \leq 1.0$. Both structural phases coexist within the $0.15 < x < 0.3$ range. Figure 2a shows the variation of lattice parameters with $x$ for both structural phases. A clear transition between Phase A and B can be identified in both the *a* and *c* lattice parameters near $x \sim 0.2$. For Phase A, both *a* and *c* change only slightly with increasing *x*, while for Phase B, *a* and *c* increase more remarkably with increasing $x$.

Phases A and B exhibit distinctly different electronic and magnetic properties. As shown in Fig. 2b and Fig. 3a, Phase A exhibits superconductivity with $T_c \sim$ 8-10 K, consistent with the reported superconductivity in FeSe$_{0.82}$ [12]. Phase B, however, exhibits enhanced superconductivity with a maximum $T_c$ of ~14 K. The normal state also displays different properties for these two phases. The temperature dependence of resistivity $\rho(T)$ for Phase A shows a metallic behavior from room temperature to the superconducting transition temperature (e.g., the data of the $x$ = 0, 0.2 samples in Fig. 3). For Phase B, however, $\rho(T)$ shows either a minimum (0.2 < $x$ < 0.6) or semiconducting-like behavior (0.6 ≤ $x$ <1.0) before the superconducting transition. The $T_c$ of Phase B varies with $x$ with the maximum ($T_c$ = 14 K) occurring near $x \sim$ 0.6. Phase B exhibits the superconducting state through $x \approx$ 0.9, but it disappears in the $x$ = 1.0 end member. The difference between the superconducting states of Phases A and B is also confirmed by magnetization measurements, as shown in Fig. 4b.

In the non-superconducting $x$ = 1.0 sample FeTe$_{0.82}$, we observed two anomalies in the magnetic susceptibility $\chi$, as denoted by arrows in Fig. 4a. One occurs near 125K, below which $\chi(T)$ exhibits a striking irreversibility between field cooling (FC) and zero-field cooling (ZFC) histories (see Fig. 4b); the other appears near 65 K where an anomalous peak in $\rho(T)$ is observed. The 125 K anomaly also occurs in all other samples with $x$ > 0.4 and this anomaly shifts down to 105-110 K when $x$ is reduced below 0.4, as shown in Fig. 4b, and Fig. 2b where the variation of the anomaly temperature $T_{ma}$ with $x$ is presented. The 65 K anomaly seen in FeTe$_{0.82}$, however, does not occur in another other sample with $x$ < 1.0.

Recent neutron scattering measurements performed by Bao *et al.* using our samples show that the 65 K anomaly in FeTe$_{0.82}$ corresponds to a simultaneous structural and antiferromagnetic transition [13], rather than the aforementioned ferrimagnetic transition [17]. The structure belongs to a tetragonal lattice with the space group *P4/nmm* at high temperatures, but distorts to a *Pmmn* orthorhombic structure below 65 K. An incommensurate antiferromagnetic order, which includes both linear and spiral components, occurs below this structure transition temperature; this magnetic order propagates along the diagonal direction of the Fe square sublattice. Such a complex magnetic behavior is different from what was observed in the parent compound of FeAs-based superconductors where the antiferromagnetic order is commensurate and propagates along one edge of the Fe square sublattice [7,8].

In addition to the antiferromagnetic transition, this structure transition also results in an anomaly in Hall coefficient. Our Hall effect measurements were performed by sweeping the magnetic field at fixed temperatures. The transverse Hall resistance $\rho_H$ exhibits a linear field dependence for each temperature. Figure 5 shows the Hall coefficient $R_H$ as a function of temperature derived from the slope of $\rho_H(H)$. We find that $R_H$ is negative and is hardly temperature dependent for $T > 65$ K, but it shows a remarkable upturn below 65 K. These observations indicate that charge carriers in FeTe$_{0.82}$ are mainly electrons and that the structure transition may lead to the change of electronic band structure and/or the variation of scattering rate of charge carries.

Regarding the magnetic anomaly near 125 K in FeTe$_{0.82}$, neutron scattering measurements did not reveal any evidence of either structure or magnetic transition [18].

Similar situations occur for other samples showing 125 K magnetic anomaly (see below). However, we note that the magnetic anomaly at 105 K in $FeSe_{0.88}$ is associated with a tetragonal-triclinic structure transformation [12], and a tetragonal-orthorhombic structural transformation at 70 K was also reported for a slightly different composition $FeSe_{0.92}$ [19]. Both results were obtained by high-resolution synchrotron x-ray diffraction measurements. Similar measurement is clearly necessary to clarify the origin of the magnetic anomalies observed in our samples.

Superconductivity in the Te/Se substituted samples appears to be related to the antiferromagnetic order in the end member $FeTe_{0.82}$. Neutron scattering measurement have been performed on the $x = 0.6$ sample which has the optimal $T_{c,max} = 14$ K. Neither long-range magnetic order nor structural transition was observed in this sample though it shows the 125 K magnetic anomaly in susceptibility (see Fig. 4b). Nevertheless short-range magnetic correlations at an incommensurate wave-vector survive and the magnetic correlation length is about 4 Å [13]. These short-range magnetic correlations depend on temperature; they start to occur below 75 K and enhance more rapidly below 40 K. Interestingly, we observed an anomalous temperature dependence in Hall coefficient in the same temperature range for this sample. As seen in Fig. 5, the Hall coefficient $R_H$ for the $x = 0.6$ sample starts to drop below 75K and a remarkable decrease occurs below 40 K, consistent with the temperature dependence of the short-range magnetic order. These observations suggest strong interplay between spin and charge degrees of freedom in this material system and that the superconducting state is extremely close to an antiferromagnetic instability. Therefore the superconductivity in the $Fe(Se_{1-x}Te_x)_{0.82}$ should be associated with magnetic fluctuations and unconventional in nature similar to other

FeAs-based superconductors [1-8]. In fact, evidence for unconventional superconductivity has been observed in recent NMR measurements for FeSe [20].

The band tuning is the most likely explanation for the presence of superconductivity in the Te/Se substituted samples. Since $Te^{2-}$ and $Se^{2-}$ have the same valence, but different ionic radii, $Te^{2-}$ substitution for $Se^{2-}$ does not directly lead to charge carrier doping, but results in the variation in band structure which in turn may change the Fermi surface. Our Hall effect measurement results shown in Fig. 5 reflect such changes. The electron density $n$ estimated from the measured Hall coefficient is $\sim 2 \times 10^{21}$ /cm$^3$ above the structure transition temperature 65K and nearly temperature independent for FeTe$_{0.82}$, while for the superconducting sample Fe(Se$_{0.4}$Te$_{0.6}$)$_{0.82}$ ($x$ =0.6), $n$ is $\sim 4 \times 10^{21}$ /cm$^3$ at 65 K and increases with increasing temperature. We note that electron densities given here for both samples are comparable to that of FeAs-based superconductors in their normal states [21].

Finally we would like to point out that the incommensurate antiferromagnetic structure of FeTe$_{0.82}$ discussed above differs from the previously-reported magnetic structure of iron telluride FeTe$_{0.90}$, which was identified as a ferromagnet for high temperatures and ferrimagnet below 63K [17]. For comparison, we also prepared a sample with the nominal composition FeTe$_{0.90}$ using the same solid-state reaction method stated above. Neutron scattering measurements on this sample show that it is truly different from FeTe$_{0.82}$ in both crystal and magnetic structure [13]. The structure transition temperature in FeTe$_{0.90}$ is shifted up to 75 K and the structure distorts to a monoclinic lattice below the transition, rather than an orthorhombic lattice as in FeTe$_{0.82}$. The antiferromagnetic order, which occurs below the structural transition, becomes commensurate, in contrast with the incommensurate antiferromagnetic order in FeTe$_{0.82}$. As noted above, the incommensurate antiferromagnetic state in FeTe$_{0.82}$ shows a semiconducting-like behavior in resistivity, while in FeTe$_{0.90}$ the commensurate antiferromagnetic state is accompanied by metallic transport properties as shown in Fig. 6. These results are inconsistent with those reported results in Ref. [17] for FeTe$_{0.90}$. One possible reason for this difference is that while our sample and the sample used in ref. [17] have the same nominal composition, their actual phase might be somewhat different since the iron telluride system has a very complicated phase diagram and the preparation conditions between our samples and the samples used in Ref. [17] are very different, which may result in subtle structural changes.

IV.    Conclusions

In summary, we report the evolution of superconductivity, magnetism, and structural transition in Fe(Se$_{1-x}$Te$_x$)$_{0.82}$ ($0 \leq x \leq 1$). The entire range of $x$ was found to be superconducting except for the $x = 1.0$ end member. Two different superconducting phases coming from two structure-types were identified, and found to coexist in the $0.15 \leq x \leq 0.3$ range. The maximum $T_c = 14$ K occurs near $x = 0.6$. In terms of the results from neutron scattering studies, the superconductivity of this system seems intimately related with magnetic correlations. The non-superconducting end member FeTe$_{0.82}$ shows an incommensurate antiferromagnetic order, while in the Te/Se substituted superconducting samples, the long-range magnetic order evolves into short-range magnetic correlations. These short-range correlations enhance significantly as the temperature is decreased below 40 K and they lead to an anomalous temperature dependence in Hall coefficient. These results strongly suggest that the superconductivity in this material system may be mediated by magnetic fluctuations, and therefore unconventional in nature. Unlike the superconductivity in high-temperature cuprates or FeAs-based compounds which is obtained through charge carrier doping, superconductivity in this system appears to be derived from band structure tuning. As a result, the Fe(Se$_{1-x}$Te$_x$)$_{0.82}$ system provides a new arena to study novel physics associated with the interplay between superconductivity and magnetism.

Acknowledgement: The work at Tulane is supported by the NSF under grant DMR-0645305, the DOE under DE-FG02-07ER46358 and the Research Corporation. Work at UNO is supported by DARPA through Grant No. HR0011-07-1-0031. Work at Penn State is supported by the DOE under Grant DE-FG02-04ER46159 and DOD ARO under Grant W911NF-07-1-0182.

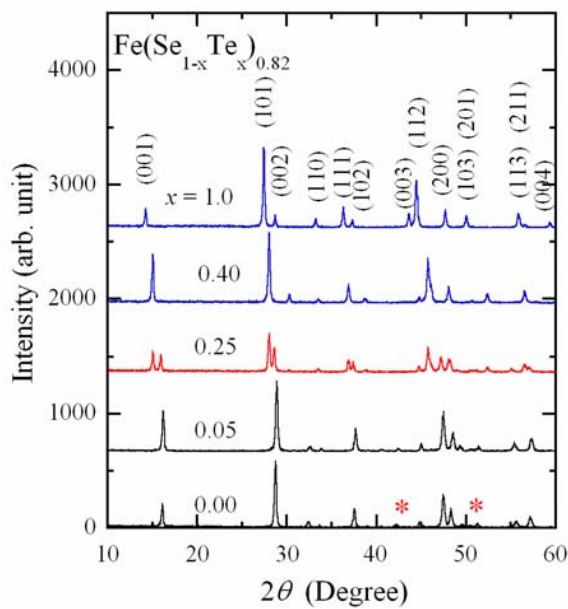

Figure 1 (Color online): X-ray diffraction patterns of typical compositions in the Fe(Se$_{1-x}$Te$_x$)$_{0.82}$ series. Two different structural phases are observed in different composition ranges. $0 \leq x < 0.15$: Phase A; $0.3 < x \leq 1$: Phase B. Phases A and B coexist in the $0.15 < x < 0.3$ range where the diffraction peaks split into two sets. While Phases A and B have the same tetragonal space group *P4/nmm*, they have remarkably different lattice parameters (see Fig. 2a). Peaks marked by * for the $x = 0$ sample: impurity phase (β-FeSe).

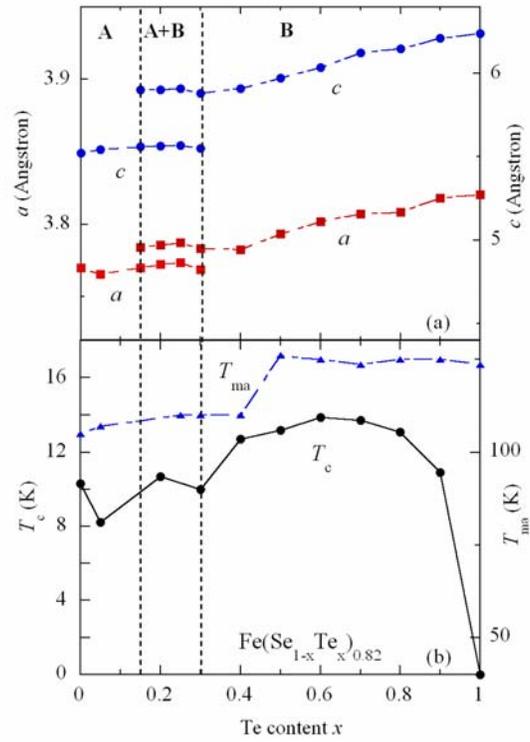

Figure 2 (Color online): Lattice parameters (a), magnetic anomaly temperature $T_{ma}$ and the onset superconducting transition temperature $T_c$ (b) as a function of Te content $x$ in the Fe(Se$_{1-x}$Te$_x$)$_{0.82}$ series. The definitions of $T_{ma}$ and $T_c$ are shown in Fig. 3 and Fig. 4 respectively.

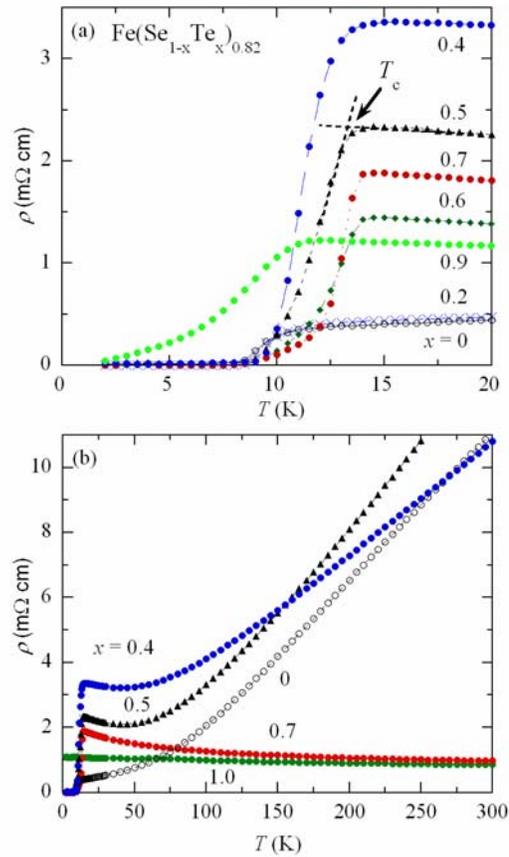

Figure 3 (Color online): Resistivity as a function of temperature $\rho(T)$ for the samples with various Te content $x$. (a) $\rho(T)$ of the samples with typical compositions for $T < 20$K. The superconducting onset transition temperature $T_c$ is defined as the intersection between the linear extrapolations of the normal state $\rho(T)$ and the middle transition, as shown in the figure. (b) $\rho(T)$ of the samples with typical compositions in the 2-300 K range.

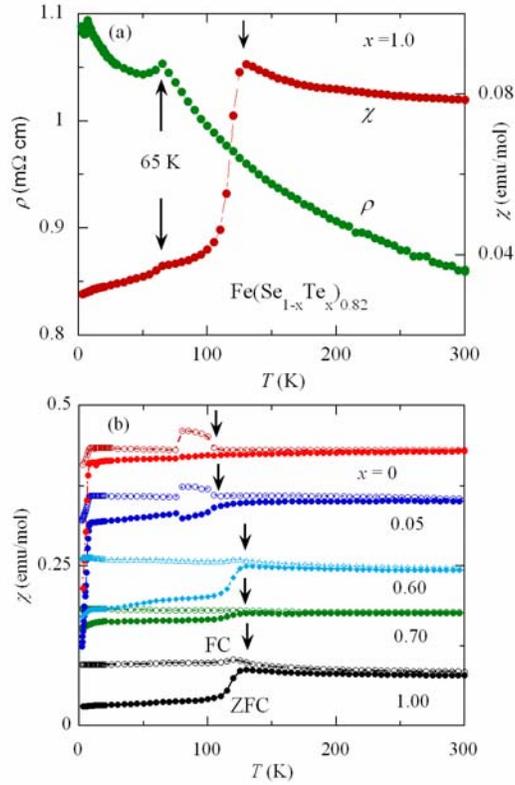

Figure 4 (Color online): (a) Magnetic susceptibility $\chi$ and resistivity $\rho$ as a function of temperature for the sample with $x = 1.0$. An anomaly near 65 K is observed in both measurements. The arrow near 125 K indicates the magnetic anomaly temperature, below which $\chi$ exhibits marked irreversibility between FC and ZFC cooling histories as shown in panel (b). (b) Magnetic susceptibility $\chi(T)$ measured following FC and ZFC cooling histories for the samples with $x = 0, 0.05, 0.6, 0.7$, and $1.0$. The transitions at low temperatures correspond to the superconducting Meissner effect. The Meissner effect is observed in all samples except for $x = 1.0$.

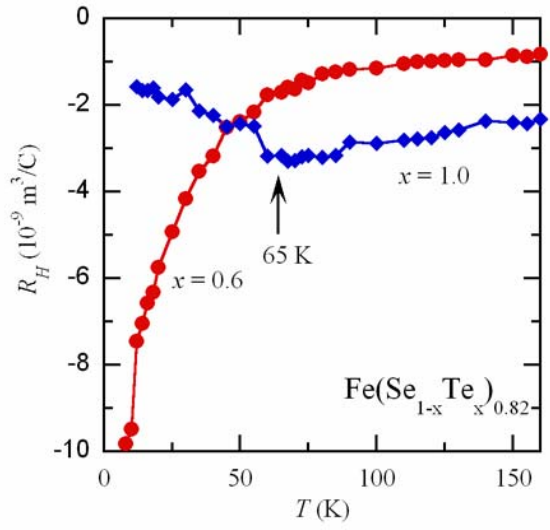

Figure 5 (Color online): Hall coefficient as a function of temperature for Fe(Se$_{1-x}$Te$_x$)$_{0.82}$ with $x = 0.6$ and $1.0$. The arrow indicates the structure transition temperature for the $x = 1.0$ sample.

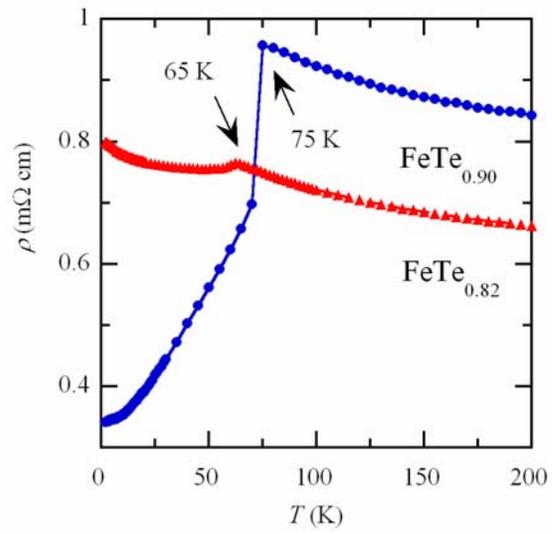

Figure 6 (Color online): Resistivity as a function of temperature for FeTe$_{0.82}$ and FeTe$_{0.90}$. The arrows indicate the structure transition temperature for each sample.